\documentclass[aps,superscriptaddress,showpacs,nofootinbib,reprint,eqsecnum,prd]{revtex4-1}

\usepackage[bf,SL,BF]{subfigure} 
\usepackage{verbatim} 
\usepackage{epsfig}  
\usepackage{amsmath}
\usepackage{amssymb}
\usepackage{textcomp}
\usepackage{amssymb}

\usepackage{slashed}

\usepackage{eso-pic}
\newcommand\AtPageUpperMyright[1]{\AtPageUpperLeft{%
 \put(\LenToUnit{0.5\paperwidth},\LenToUnit{-1cm}){%
     \parbox{0.5\textwidth}{\raggedleft\fontsize{9}{11}\selectfont #1}}%
 }}%
\newcommand{\conf}[1]{%
\AddToShipoutPictureBG*{%
\AtPageUpperMyright{#1}
}
}

\input{epsf}
\def\be{\begin{equation}} 
\def\ee{\end{equation}} 
\def\bea{\begin{eqnarray}}
\def\eea{\end{eqnarray}}

\def\beq{\begin{equation}}
\def\eeq{\end{equation}}
\def\beqa{\begin{eqnarray}}
\def\eeqa{\end{eqnarray}}

\newcommand{\bs}{\begin{split}}
\newcommand{\es}{\end{split}}

\newcommand{\dfp}{\frac{d^4p}{(2\pi)^4}}

\newcommand{\trd}{{\rm \,tr}}
\newcommand{\eq}[1]{Eq.~(\ref{#1})}

\newcommand{\fig}[1]{Fig.~\ref{#1}}

\begin{document}

\title{Optimized Perturbation Theory Applied to a
Model with Flavour Symmetry $SU_f(3)$  }

 \author{Juan C. Mac\'ias}  \email{jcamilo@if.usp.br}
\affiliation{Instituto de  de F\'{\i}sica, Universidade  de S\~ao Paulo,
C.P. 66318, 05315-970 S\~ao Paulo, SP, Brazil}    
 
\author{Marcus Benghi Pinto}  
 \email{marcus.benghi@ufsc.br} 
\affiliation{Departamento de F\'{\i}sica, Universidade Federal de Santa
  Catarina, 88040-900 Florian\'{o}polis, Santa Catarina, Brazil}

\begin{abstract}
The \textit{optimized perturbation theory} (OPT) is implemented in the $SU_f(3)$ flavor symmetric Nambu--Jona-Lasinio (NJL) model to generate non-pertubative corrections to the quark pressure beyond the large-$N$ approximation. The correctness of this implementation is verified by the recovery of the already known non-perturbative results in the Hartree-Fock approximation, and by having the large-$N$ approximation as a  limiting case. This formalism is then used to revisit a discussion on the discordance between the lattice data and the two flavor model prediction of the dynamical vector repulsive interactions, beyond the pseudocritical temperature. It is shown that these contradictory predictions can be corrected by considering a three quark flavor system. 

\end{abstract}

\maketitle
\conf{ XIV Hadron Physics, Florianópolis, Brazil, March 18-23 of 2018.}
\section{Introduction}%
\label{Sec:Intro}

The \textit{optimized perturbation theory} (OPT) also known as $\delta-$expansion \cite{duncan1988nonperturbative, okopinska1987nonstandard} is a theoretical scheme created to study non-perturbative aspects of the quantum field theories. By combining perturbative results with a variational criterion, it provides a method to perform nonperturbative calculations beyond the leading order of approximation in many contexts and applications. One of such applications was presented by us in  Ref.\cite{Restrepo:2014fna},  where  corrections to the quark pressure and some related quantities, were calculated using the OPT  in the $SU_{f}(2)$ flavor symmetric Polyakov--Nambu--Jona-Lasinio (PNJL) model \cite{reg}. The dynamically generated finite-$N_c$ corrections ($N_c$ being the number of color charges), consisted in repulsive vectorial terms,  similar to those found in model calculations restricted to the large-$N_c$ (LN) approximation when an explicit vector channel is added to the Lagrangian. In the OPT case, the repulsive vector contributions being generated in a dynamic way, end up having an intensity proportional to $G_S/N_c$, where $G_S$ is a parameter already present in the model as the scalar coupling constant. In the LN case, the intensity of the vector repulsive terms are regulated by the additional parameter $G_V$, interpreted as a vector coupling constant. These vector repulsive terms,  are responsible for a better description of the lattice data in temperatures below the pseudocritical temperature $T_c$, but spoil that description above $T_c$. Specifically, the second coeficient $c_2$, of the Taylor expansion of the pressure around zero chemical potential: $P/T = c_0 + c_{2}(\mu/T)^2+\cdots$, does not converge to the Stefan-Boltzmann limit and develops a maximum at $T\sim 1.2\,T_c$,  not registered by any lattice simulation.


The discrepancy between the model with an explicit vector interaction and the lattice results for $c_2$ above $T_c$, led the authors in Ref. \cite{stefanGV} to the conclusion that strong vector interactions can only be  present in the hadronic phase they being essentially null in the deconfined phase where $G_V$ must be set to zero. 

Although our model calculation basically supported that conclusion, given that in our model the vector repulsive terms are parametrized by $G_S$ and not by $G_V$,  we showed that the same conclusion can be achieved even when $G_V =0$ in both regimes.

One possible solution to this problem, could be that the vector contributions be naturally eliminated at high $T$ if $G_S$ goes to zero as higher order contributions are calculated.

In this conference proceeding, we report on one alternative possibility. We will show that the dynamically induced repulsive vector interactions, present in the chiral symmetric phase, can be a direct consequence of calculating finite-$N_c$ corrections in a  model that only considers two light quarks. To that end, we will implement the OPT in the Nambu--Jona-Lasinio (NJL) model with flavor symmetry $SU_f(3)$ which will allow us to study the physical consequences of treating the OPT with a third light quark in the system. 

This work is organized as follows: in Sec. \ref{Sec:model} we present our implementation of OPT in the  $SU_{f}(3)$ flavor symmetric  NJL model. The validity of our implementation  prescription, is confirmed by showing that it can generate in an alternative manner the already known nonperturbative results of the Hartree-Fock (HF) approximation. We also verify that the Large-$N_c$ approximation is recovered as the limiting case $N_c \rightarrow \infty$. We present our numerical results for $c_2$ in Sec. \ref{Sec:results} and  our conclusions in Sec. \ref{Sec:conclusions}

\section{Model framework}
\label{Sec:model}
The basic idea behind the OPT consists in replacing the original Lagrangian of the theory by one containing an arbitrary parameter $\eta$. A Gaussian term proportional to $\eta$, that hence does not modify the dynamics,  is added to the Lagrangian, while the same term multiplied by the fictitious expansion parameter $\delta$ is subtracted.  Given a theory that is described by a Lagrangian density $\mathcal{L}$, a new interpolated Lagrangian $\mathcal{L}^{\delta}$ is defined such that
\begin{align}
\mathcal{L}^{\delta} = (1 -\delta)\mathcal{L}_{0}(\eta) + \delta\mathcal{L}.
\end{align}
 $\mathcal{L}^{\delta}$ interpolates between a solvable theory $\mathcal{L}_{0}$ (when $\delta = 0$), and the original theory $\mathcal{L}$ (when $\delta = 1$).  The parameter $\delta$ is also used as a dummy label to index the order of the perturbative calculations and for that purpose all the couplings of the original theory are multiplied by $\delta$ . All pertubative calculations are performed in powers of $\delta$, which is formally treated as as being small and fixed to its original value $\delta = 1$ at the end of calculations. Meanwhile, the $\eta$ parameter can be viewed as a mass parameter and since the fermionic propagator gets dressed by $\eta$, it also serves as a infrared regulator.

Although these modifications do not change the dynamics  defined by the original Lagrangian, the Feynman rules are different. The physical quantities of interest, can now be evaluated up to the order $\delta^k$ using the new (but trivial), Feyman rules of the modified theory. Once a physical quantity $P$ is evaluated to order $\delta^k$ and $\delta$ is fixed to unity, a residual dependence on $\eta$ remains. Nonperturbative results can be obtained requiring that $P(\eta)$ be considered in the point where it is less sensitive to variations. This can be achieved by using the variational criteria known as the \textit{Principle of Minimal Sensibility}  (PMS)\cite{pms} enunciated as  
\begin{align}
 \frac{dP(\eta)}{d\eta }\Bigg\vert _{\bar{\eta},\delta
  =1}=0\;.
\label{pms12}
\end{align}
For this application, we are going to use the three flavour version of the NJL model. The NJL model can be de described by the Lagrangian density
\begin{eqnarray}
\begin{split}
\mathcal{L} = \bar{\psi}(i\slashed{\partial} &- \hat{m}_f)\psi + G \sum_{a = 0}^8\left[
(\bar{\psi}\lambda^a \psi)^2 + (\bar{\psi}i\gamma_5\lambda^a \psi)^2\right] \\
&- K \left[{\rm det}\,\bar{\psi}(1+\gamma_5 )\psi + {\rm det}\,\bar{\psi}(1-\gamma_5 )\psi
\right],
\end{split}
\label{Eq:LagrangianaSU(3)}
\end{eqnarray}
where $\psi = (u,d,s)^{T}$ represents a quark field with three flavors, $\hat{m}_f = {\rm diag}(m_u,m_d, m_s)$ is the corresponding (current) mass matrix, $\lambda_0 = 2/3 I$ where $I$ is the unit matrix in the
three flavor space, and $0 < \lambda_a \leq 8$ denote the Gell-Mann
matrices. We consider $m_u = m_d \neq m_s $. The terms proportional to the constants $G$ and $K$ describe local 4-fermion and 6-fermion interactions between quark fields $\psi$, respectively.  In the 4-fermion interaction vertex, only quarks fields with the same flavor charge intervene, while the 6-fermion term mixes quarks flavors due to the determinant in  the flavor space.
\begin{figure}
\includegraphics[width=.6\linewidth]{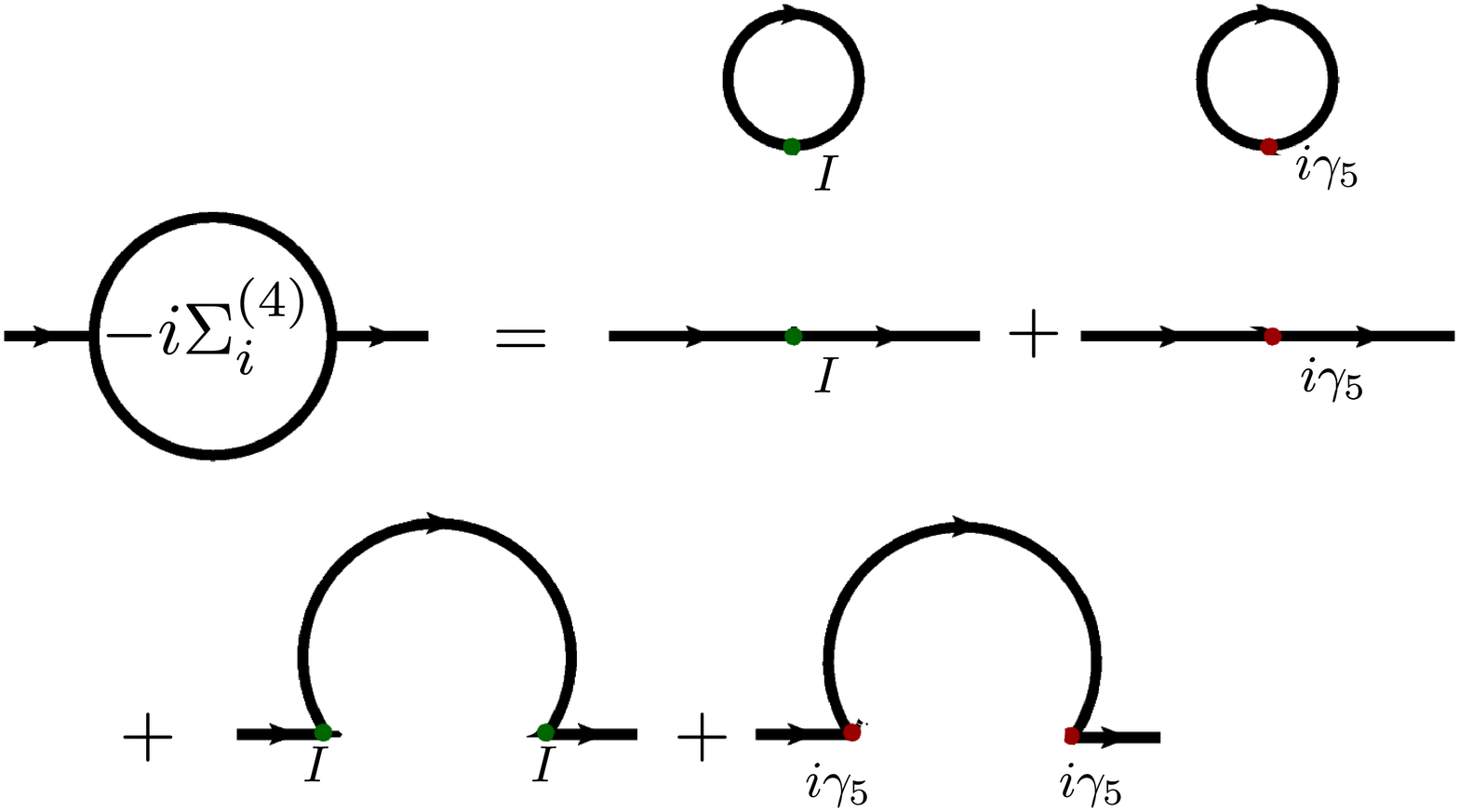}  
\includegraphics[width=.6\linewidth]{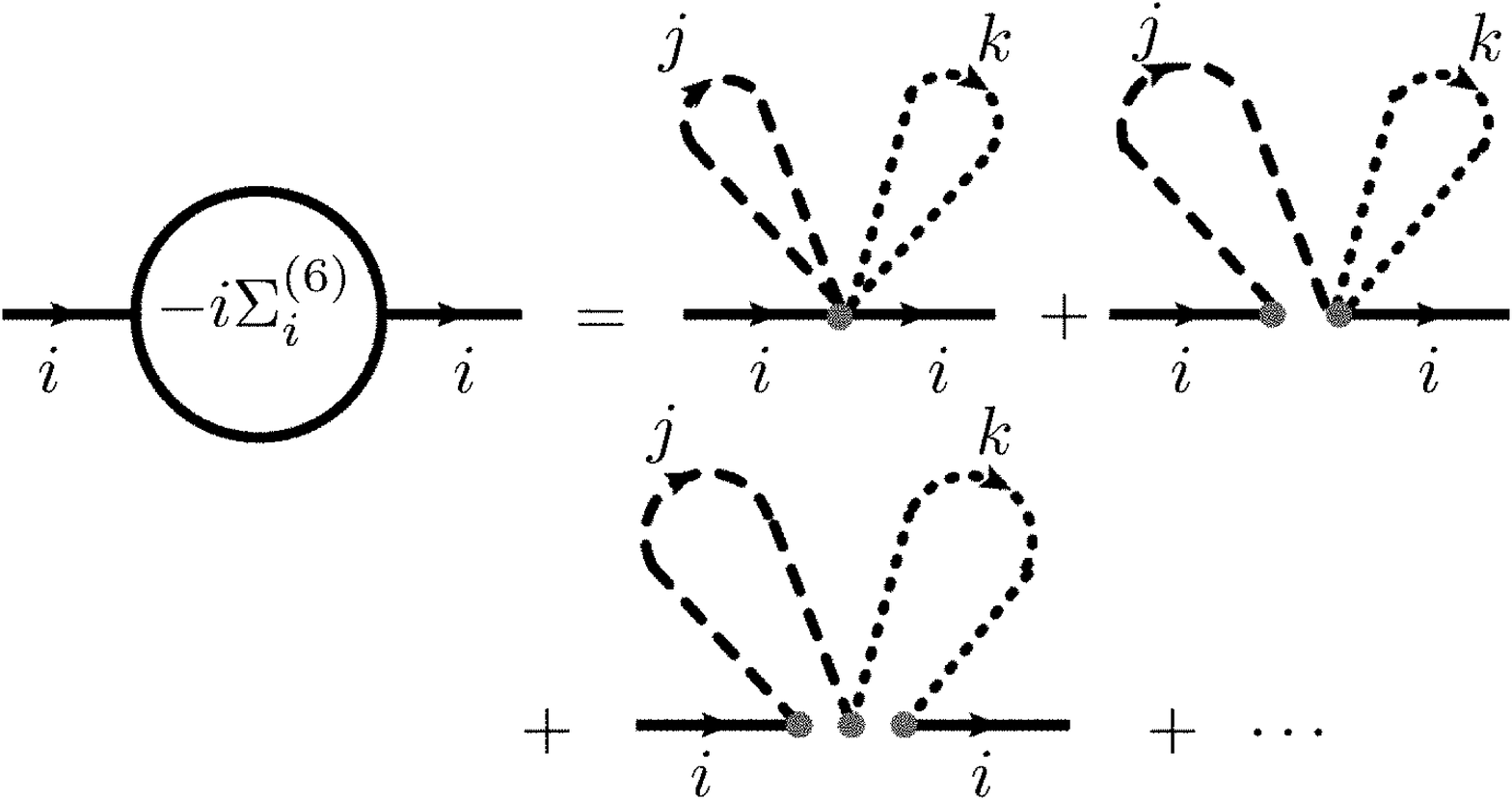}  
\caption{Hatree and Fock contributions to selfenergy. (Top) Self energy associated to the 4-fermion interaction. (Bottom)Self energy diagrams associated to the 6-fermion interaction. }
  \label{fig:Self4&6}
\end{figure}    
Considering the Hartree and Fock contributions to the selfenergy in \fig{fig:Self4&6}, it is possible to see that the effective quark masses  are related through the set of selfconsistent relations
\begin{align}
\bs
M_{i}&=  m_i -4G_{s}\phi_{i} + 2K\Big(1+ \frac{3}{2N_c} +\frac{1}{2N_{c}^{2}}\Big)\phi_{j}\phi_{k}\\
&=m_i -4G_{s}\phi_{i} + 2K^{\prime}\phi_{j}\phi_{k}  \qquad(i\neq j\neq k),
\label{Eq:GapSU3}
\es
\end{align}
also known as gap equations. Where $K^{\prime} = K(1+ 3/2N_c +1/2N_{c}^{2})$. The \eq{Eq:GapSU3}, gives the numerical values of the effective masses of the considered quark flavors $i = u, d, s$, in terms of which the Landau free energy (effective potential) of the model reads \cite{Klevansky:1992qe}
\begin{align}
\bs
\mathcal{F} = -\theta_u - \theta_d  -\theta_s +2G(\phi_{u}^{2}+\phi_{d}^{2}&+\phi_{s}^{2} )\\
& - 4K^{\prime}\phi_{u}\phi_{d}\phi_{s}.
\es
\label{Eq:E_LivreHFSU3}
\end{align}
Where $\theta_i$ and $\phi_{i}$ are respectively the quasiparticle and scalar density contributions to the free energy associated to the quark with flavor $i$, given as 
\begin{align}
\theta_{i} = -iN_{c}\trd\int \dfp \ln(\slashed{p} - M_i) 
\end{align}
and 
\begin{align}
\phi_{i} = -iN_c\trd\int \dfp \frac{1}{\slashed{p} - M_i}
\end{align}
respectively. Here ``tr'' represents the trace on the Dirac space.

The non-perturbative nature of Eqs. \eqref{Eq:GapSU3} and \eqref{Eq:E_LivreHFSU3} is clear from the dependence on the effective mass $M_i$, which is nonperturbatively generated by the dressing of the bare quark mass $m_i$  through the resuming of the infinite selfinteractions of the Hartree and Fock type. Nonetheless, we could also sum these Hartree and Fock contributions in a perturabative fashion in powers of the coupling constants $G$ and $K^{\prime}$. It would result in the expression for the free given as
\begin{align}
\begin{split}
\mathcal{F}^{\rm PT}= -\theta_{0u}  &-\theta_{0d}  -\theta_{0s}  - 2G(\phi_{0u}^{2} +\phi_{0d}^{2} +\phi_{0s}^{2})\\
& +   2K^{\prime}\phi_{0u}\phi_{0d}\phi_{0s} + \mathcal{O}(G^{2}, K^{\prime 2}),
\end{split}
\label{Eq:Pressao_PT}
\end{align}
which, unlike \eq{Eq:E_LivreHFSU3}, does not depend directly on effective mass $M_i$, but  in the current mass $m_i$ through the quasiparticle and scalar density terms $\theta_{0 i}$ and $\phi_{0 i}$, given as
 \begin{align}
\theta_{0i} = -iN_{c}\trd\int \dfp \ln(\slashed{p} - m_i) 
\end{align}
and 
\begin{align}
\phi_{0i} = -iN_c\trd\int \dfp \frac{1}{\slashed{p} - m_i}
\end{align}
respectively.

The perturbative form of the free energy $\mathcal{F}^{\rm  PT}$ in \eq{Eq:Pressao_PT} is suitable to the application of the OPT formalism.

We want now to illustrate how the nonperturbative results, in Eqs. \eqref{Eq:GapSU3} and \eqref{Eq:E_LivreHFSU3}, can be also deduced from the perturbative expression in \eq{Eq:Pressao_PT} by the application of the OPT prescriptions.

To use the OPT on this model, we are gonna use the following interpolation prescription:
\begin{enumerate}
\item Interpolate in the quark masses by adding and subtracting a parameter $\eta_i$ for each quark flavor $i$, doing
\begin{align}
m_i \rightarrow m_i + (1-\delta)\eta_i.
\label{Eq:massInterpolation}
\end{align}
\item Multiply all the vertices by the expansion parameter $\delta$, such that
\begin{align}
G, \,K^{\prime} \rightarrow \delta G,\, \delta K^{\prime} .
\label{Eq:VertexOPT}
\end{align}
\end{enumerate} 
According to this prescription, the interpolated  free energy up to  order $\delta$ is given by
\begin{align}
\begin{split}
\mathcal{F}^{\rm OPT}_{\delta^1} &=- \theta_{u} - \theta_{d} - \theta_{s} -\delta(\eta_{u}\phi_{u}
+\eta_{d}\phi_{d}+\eta_{s}\phi_{s}) \\
&- 2\delta G(\phi_{u}^{2}
+\phi_{d}^{2}+\phi_{s}^{2}) + 2\delta K^{\prime}\phi_{u}\phi_{d}\phi_{s} .
\end{split}
\end{align} 
Then, by applying the PMS to the interpolated free energy we get 
\begin{align}
\begin{split}
&\frac{\partial  \mathcal{F}^{\rm OPT}_{\delta^1}}{\partial \eta_i }\bigg\rvert_{\eta_i = \bar{\eta_i }, \delta = 1} = 0\\
&= \phi_{i} -\bar{\eta_{i}}\frac{\partial \phi_{i}}{\partial \eta_{i}}\bigg\vert_{\bar{\eta_i}} - \phi_{i} -4G\phi_{i}\frac{\partial \phi_{i}}{\partial \eta_{i}}\bigg\vert_{\bar{\eta_i}} + 2K^{\prime}\frac{\partial \phi_{i}}{\partial \eta_{i}}\bigg\vert_{\bar{\eta_i}}\phi_{j}\phi_{k}.
\end{split}
\label{Eq:PMS_SU3}
\end{align}
This equation is satisfied if
\begin{align}
\bar{\eta_{i}} =- 4G\phi_{i} + 2K^{\prime}\phi_{j}\phi_{k}.
\label{Eq:GapOPT_SU3}
\end{align}
So, identifying $\bar{\eta_{i}}$ with the effective mas $M_i$, such that $\bar{\eta_{i}} = M_i - m_i$, we verify that \eq{Eq:GapOPT_SU3} its equivalent to the gap equations \eq{Eq:GapSU3}. Likewise, by taking this result into the perturbative version of the free energy, \eq{Eq:Pressao_PT}, we recover the non-perturbative result  in \eq{Eq:E_LivreHFSU3}, it is
\begin{align}
\begin{split}
&\mathcal{F}^{\rm OPT}\bigg\vert_{\eta_i = \bar{\eta_i}, \, \delta = 1} =- \theta_{u} - \theta_{d} - \theta_{s} -(- 4G\phi_{u} \\
&+ 2K^{\prime}\phi_{d}\phi_{s})\phi_{u}-(- 4G\phi_{d} + 2K^{\prime}\phi_{i}\phi_{s})\phi_{d}\\
&-(- 4G\phi_{s} + 2K^{\prime}\phi_{i}\phi_{d})\phi_{s}) - 2 G(\phi_{u}^{2}+\phi_{d}^{2}+\phi_{s}^{2})\\
& + 2 K^{\prime}\phi_{u}\phi_{d}\phi_{s} \\
& = - \theta_{u} - \theta_{d} - \theta_{s} + 2 G(\phi_{u}^{2}+\phi_{d}^{2}+\phi_{s}^{2})\\
&\qquad\qquad\qquad\qquad\qquad\qquad- 4 K^{\prime}\phi_{u}\phi_{d}\phi_{s}.
\end{split}
\label{Eq:F_OPTSU3}
\end{align}
Which is indeed the same expression given by \eq{Eq:E_LivreHFSU3}. 

The fact that we are able to generate through  the application of the  particular OPT prescription in Eqs. \eqref{Eq:massInterpolation} and  \eqref{Eq:VertexOPT}, the already known non-perturative results, serves as a crosscheck for the interpolation strategy used here. We note  that, when $N_{c}\rightarrow \infty$ also $K^{\prime} \rightarrow K$  and the large-$N_{c}$ results are recovered.

\section{Results and discussion}
\label{Sec:results}
\begin{figure}
\includegraphics[width=.9\linewidth]{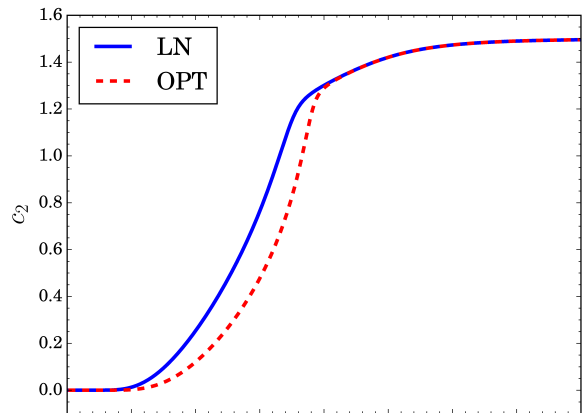}  
\includegraphics[width=.85\linewidth]{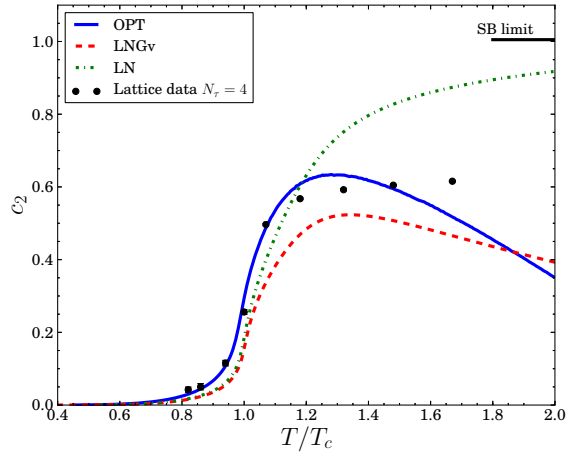}  
\caption{Second order coefficient $c_2$ of the  Taylor expansion of pressure along zero chemical potential. Top: $c_2$ against $T$ in the $SU_f(3)$ symmetric NJL model in the OPT and LN approximation. Bottom: Adapted from \cite{Restrepo:2014fna}. $c_2$ against $T/T_c$ in the $SU_f(2)$ symmetric PNJL model in the OPT and LN approximation and LN with vector interaction (LNGv) }
  \label{fig:c2}
\end{figure}   
In the bottom  panel  of \fig{fig:c2} we include, for comparison, our previous result for  $c_2$ in the $SU_f(2)$ flavor symmetric PNJL model in Ref. \cite{Restrepo:2014fna}. As stated in the introduction, when the OPT is considered in this case, it is observed  a maximum in $c_2$ , not described by the lattice data,  at $T\sim 1.2 \, T$, similar to the maximum obtained in the LN approximation when vector channel is considered.  

In the top panel of \fig{fig:c2} it is plotted  the second order coefficient of the Taylor expansion of the pressure along zero chemical potential, 
obtained  by using the  OPT  and the LN approximation in the $SU_{f}(3)$ flavor symmertric NJL model. In this case, the $c_2$ coefficient recovers the convergence to the  Stefan-Boltzmann limit. The main difference here with respect the two flavor case, is  that despite the fact that also in this case, the quark pressure gets Fock repulsive vector-like contributions, those contributions end up canceling each other exactly. This remarkable observation,  can be better understood by comparing the vertex structures of the two models. To that end, let us rewrite the usual two flavor interaction term of the NJL model Lagrangian, in a form more suitable for comparison with its three flavor counterpart, such that
\begin{align}
\bs
{\cal L}_{\rm int,2} &=G_{S} \Big[(\bar{\psi}\psi)^{2} + (\bar{\psi}i\gamma_{5}\vec{\tau}\psi)^{2}\Big]\\ & = {\cal L}_{\rm sym,2} + {\cal L}_{\rm det,2},
\es
\end{align} 
where
\begin{align}
{\cal L}_{\rm sym,2} = \frac{1}{2} G_{S}\sum_{i=0}^{3} \Big[(\bar{\psi}\tau^{i}\psi)^{2} + (\bar{\psi}i\gamma_{5}\tau^{i}\psi)^{2}\Big]
\label{Eq:Lsym}
\end{align}
and
\begin{align}
\bs
{\cal L}_{\rm det,2} &= \frac{1}{2} G_{S} \Big[(\bar{\psi}\psi)^{2} + (\bar{\psi}i\gamma_{5}\vec{\tau}\psi)^{2}\\
&\qquad\qquad - (\bar{\psi}i\gamma_{5}\psi)^{2}-(\bar{\psi}\vec{\tau}\psi)^{2}\Big]\\
& = G_{S}\Big[{\rm det}\,\bar{\psi}(1+\gamma_{5})\psi +{\rm det}\, \bar{\psi}(1-\gamma_{5})\psi\Big].
\es
\label{Eq:Ldet2}
\end{align}
It can be verified  by a simple analysis, that all the finite $N_c$ corrections come from the determinantal terms in Eqs. \eqref{Eq:Ldet2} and \eqref{Eq:LagrangianaSU(3)}. The structure of this term in \eq{Eq:Ldet2}, corresponds to a 4-fermion interaction, ie, it describes the interaction of two bodies (two quark flavors). On the other hand, the corresponding term in the three flavor model in \eq{Eq:LagrangianaSU(3)}, is a 6-fermion interaction, or in other words, describes a three body interaction. The differences on the nature  of the interactions, allow the the Fock corrections associated to $\mathcal{L}_{\rm det,2}$ to be vectorial and repulsive, while the same terms associated to the 6-fermions interaction of the three flavor model, are canceled by counterparts of opposite signs.

\begin{figure}
\includegraphics[width=.85\linewidth]{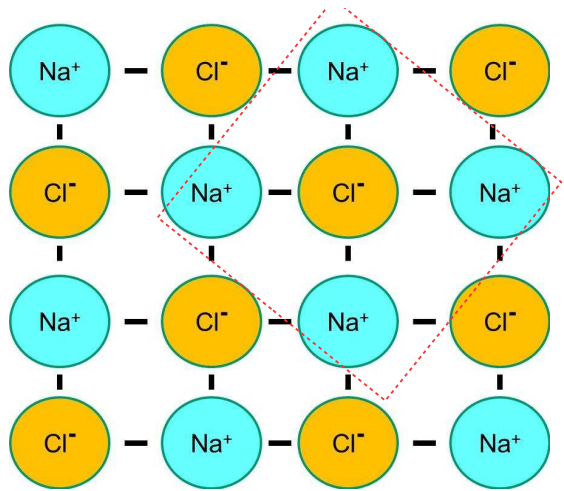}  
\caption{\small  Salt crystal. In a salt
crystal, all the sites in the lattice are occupied by chloride atoms with negative net electric charge and sodium
atoms with positive net electric charge.}
  \label{fig:Salt}
\end{figure}

The fact that in a system described by two body interactions, the finite $N_c$ correction  manifest themselves as repulsive in character, while in a system described by three body interactions these repulsive interactions be canceled, should not come as a surprise. Let us consider for example the classical analogy presented in \fig{fig:Salt}, where a salt crystal  
is depicted. In a salt crystal, all the sites in the lattice are occupied by chloride atoms with negative net electric charge and sodium atoms with positive net electric charge. One way to calculate the electrical  force at a certain distance from the crystal would be to idealize this system as and homogeneous and neutral charge distribution as a consequence of considering the number of atoms in the lattice as a large number (``large $N$ aproximation''). A better idealization, would consider the fact that the crystal is a composite object with a finite number of atoms. ``Finite $N$ corrections'' to the electric force can be calculated then, by either considering the two different kind of atoms present (sodium and chloride) or by only considering one kind of atom. For example, if our test charge is near  the region enclosed by the dashed line in \fig{fig:Salt}, we could  consider only the effects of sodium atoms. In the former case, the calculated finite $N$ corrections to the electrical force will be repulsive in character since all the atoms considered have the same electrical charge. I the later case, the repulsive character of the finite $N$ corrections to the electrical force, is going to be attenuated by the inclusion of atoms with opposite electric charge. 

In a analogous way, when finite $N_c$ corrections to the quark pressure are considered in a system approximated by one containing only two light quarks, the finite-$N_c$ corrections manifest themselves as vector repulsive in character. But if those finite-$N_c$  corrections are calculated in a system described in terms of three light quarks, it is reasonable to expect that  those vector repulsive contributions be attenuated or as in this particular case, completely canceled.
\section{Conclusions}
\label{Sec:conclusions}  
We have implemented the OPT in the  $SU_f(3)$ flavor symmetric NJL model. Our implementation proves to be very simple and straightforward, basically we add and subtract a mass parameter for each quark flavor and keep track of the perturbative orders by multiplying all vertexes by the expansion parameter $\delta$. However, although  there is no a unique  way to implement the  OPT on this model, the simple exercise presented here, is validated by the recovery of the already known non-pertubative calulations of the Hartree-Fock approximation and also by having the LN result as the limit case where $N_c \rightarrow \infty$. 

We have also revisited the discrepancies between the lattice data of the second order coefficient of the Taylor expansion of pressure at zero chemical potential and the results from the OPT in the two flavor PNJL  model or the same model with an explicit vector channel in the LN approximation. We have discussed  how the inclusion of a vector channel or a dynamically generated repulsion is important for a more realistic description of the hadronic phase. The perturbative behavior expected at hight $T$, can only be described by the model with two quark flavors if the vector interactions are null in that regime. We have shown that this happens naturally when three quark flavors are considered and the vector repulsive terms are dynamically generated. On the other hand, this study suggests, that if the a vector channel is explicitelly included in the model, the vector coupling constant  must be set to zero in the deconfined phase as argued in Refs. \cite{stefanGV, stefanGV2}.

\section*{Acknowledgements} 
This work was partially supported by  CAPES (Brazil) and by CNPq
(Brazil) under process No. 171116/2017-8.

\bibliographystyle{apsrev4-1} 
\bibliography{xampl} 

\end{document}